\documentclass[%
 reprint,
 amsmath,amssymb,
 aps,
]{revtex4-2}
\usepackage{graphicx} 

\usepackage{amsmath,amssymb,amsfonts,comment}
\usepackage{slashed,color,xcolor,url,hyperref}
\usepackage[margin=2cm]{geometry}
\usepackage[normalem]{ulem}

\newcommand{\bra}[1]{{\big\langle #1 \big\vert}}
\newcommand{\ket}[1]{{\big\vert #1 \big\rangle}}
\newcommand{\braket}[2]{{\big\langle #1 \big\vert #2 \big\rangle}}

\newcommand{\T}[1]{\boldsymbol{#1}}

\begin{document}
\title{Relativistic corrections and high-energy resummation \\ for exclusive heavy quarkonium photoproduction}

\author{Maxim Nefedov}
 \altaffiliation{Physics Department, Ben-Gurion University of the Negev, \\
Beer Sheva 84105, Israel}
 \email{nefedov@post.bgu.ac.il}

\begin{abstract}
  The $O(v^2)$ correction to the high-energy resummed coefficient function for the exclusive photoproduction of vector ($1^{--}$) heavy quarkonia off hadrons is computed, thereby taking into account corrections of $O(v^2 \alpha_s^n \ln^{n-1}(x/\xi))$. When expressed in terms of the physical vector-meson mass ($M_V$) and the relative heavy-quark velocity ($\langle v^2 \rangle$) using the Gremm--Kapustin relation, the computed correction turns out to be negligible at the scale $\mu_F=M_V$. However, it partially cancels the $\mu_F$ dependence of the $O(v^2)$ correction to the LO coefficient function in $\alpha_s$, thereby improving the robustness of the predictions. 
\end{abstract}

\maketitle

\section{Introduction}

The exclusive photoproduction of (pseudo-)vector ($1^{--}$) charmonia and bottomonia is one of the most direct channels for probing the gluon distributions of the proton and nuclei~\cite{Ryskin:1992ui,Jones:2015nna,Jones:2016ldq,Flett:2019pux,Guzey:2013xba,Guzey:2013qza,Eskola:2023oos,Eskola:2022vaf,Eskola:2022vpi} (see also Ref.~\cite{Boer:2024ylx} for an EIC-centric review). The experimental study of this process at highest photon-hadron collision energies is possible with the help of ultra-peripheral heavy-ion collisions at the LHC~\cite{Bertulani:2005ru, Baltz:2007kq, Contreras:2015dqa, Klein:2019qfb}. Within the framework of collinear factorisation (CF), the coefficient function for this process has been known up to NLO in $\alpha_s$ for a long time~\cite{Ivanov:2004vd}. Very recently, the first relativistic $O(v^2)$ correction to the LO coefficient function in $\alpha_s$ has also been computed~\cite{Blask:2025jua} within the framework of Non-Relativistic QCD (NRQCD) factorisation~\cite{Bodwin:1994jh}.

At high photon-hadron collision energies, $\sqrt{s}\gg M_V$, low-$x$ QCD effects become important. The first manifestation of this was already noticed in Ref.~\cite{Ivanov:2004vd} and in subsequent phenomenological studies~\cite{Jones:2016ldq,Eskola:2023oos,Eskola:2022vaf,Eskola:2022vpi,Flett:2019pux}. It was found that the NLO amplitude develops a strong dependence on the factorisation scale $\mu_F$, which undermines the predictive power of the collinear-factorisation calculation at high energies~\footnote{Similar effects in inclusive heavy-quarkonium production have been studied in Refs.~\cite{Lansberg:2021vie,Lansberg:2023kzf}.}. A possible remedy for this problem would be to switch entirely to a factorisation formalism designed for the low-$x$ regime, such as the colour-glass condensate (CGC), where both NLO corrections in $\alpha_s$ and $O(v^2)$ corrections are currently available~\cite{Mantysaari:2022kdm}. However, the factorisation of the amplitude at low $x$ has a structure that is fundamentally different from that of CF, making it difficult to connect these two descriptions (see, however, the recent Ref.~\cite{Bhattacharya:2025fnz} for progress in this direction). Moreover, the leading eikonal approximation of the CGC formalism is specifically tailored to the asymptotic regime $\sqrt{s}\gg M_V$. Strictly speaking, it is therefore unclear at which energies the CF description should be abandoned in favour of the asymptotic high-energy description.

An alternative approach to this problem was proposed within CF in Refs.~\cite{Ivanov:2007je,Ivanov:2015hca}. The high-energy instability of the $\mu_F$ dependence of the NLO amplitude, first identified in Ref.~\cite{Ivanov:2004vd}, arises because the region $\xi\ll |x| \ll 1$ begins to dominate the collinear convolution integral at NLO in $\alpha_s$, owing to the fact that the ``skewness parameter'' satisfies $\xi \sim M_V^2/(2s)\ll 1$. Perturbative resummation of the series of corrections $O(\alpha_s^n \ln^{n-1}(x/\xi))$ (the {\it Leading Logarithmic Approximation}, LLA) to the CF coefficient function using the {\it High-Energy Factorisation} (HEF) formalism of Refs.~\cite{Collins:1991ty,Catani:1990xk,Catani:1990eg} cures this instability and partially cancels the $\mu_F$ dependence of the LO amplitude in $\alpha_s$, thereby improving the predictive power of the calculation. In Ref.~\cite{Flett:2024htj}, the formalism of Ref.~\cite{Ivanov:2007je} was adapted to computations involving Generalised Parton Distributions (GPDs), whose evolution is governed by fixed-order (LO or NLO) evolution kernels\footnote{In principle, consistency with the complete LLA computation of the coefficient function would require the GPDs to evolve with $\mu_F$ according to low-$x$ LLA-resummed evolution kernels. The DLA resolves this inconsistency, allowing one to use the conventional GPDs extracted from global fits.}. To this end, the {\it Doubly-Logarithmic Approximation} (DLA) of HEF~\cite{Lansberg:2021vie} was applied to exclusive photoproduction in Ref.~\cite{Flett:2024htj}.

In the present paper, the $O(v^2)$ correction to the DLA HEF-resummed coefficient function of Ref.~\cite{Flett:2024htj} is computed using the NRQCD factorisation formalism. At high energies, this $O(v^2 \alpha_s^n \ln^{n-1}(x/\xi))$ correction should be treated on the same footing as the $O(v^2)$ correction to the LO coefficient function computed in Ref.~\cite{Blask:2025jua}, since the resummation parameter at the amplitude level (after convolution with GPD) satisfies $\alpha_s \ln(1/\xi) \sim 1$. However, when expressed in terms of the physical quarkonium mass $M_V$ and the parameter $\langle v^2\rangle$ using the Gremm--Kapustin relation of NRQCD~\cite{Gremm:1997dq}, the correction computed in the present paper is found to be numerically small at the scale $\mu_F=M_V$ compared with the $O(v^2)$ correction of Ref.~\cite{Blask:2025jua}. It should be noted, however, that the $\mu_F$ dependence of the $O(v^2)$ correction to the LO amplitude is partially cancelled by the $O(v^2)$ correction to the HEF-resummed amplitude, thereby reducing the scale dependence of the cross section, so it makes sense to include the latter one into the phenomenological analysis.

The remainder of this paper is organised as follows. In Sec.~\ref{sec:NRQCD}, we discuss the general aspects of the NRQCD matching procedure and different ways of representing the $O(v^2)$-corrected amplitude. To establish our notation, Sec.~\ref{sec:LO-CF-v2-corr} reviews the $O(v^0)$ coefficient function in CF together with the $O(v^2)$ correction to the LO coefficient function computed in Ref.~\cite{Blask:2025jua}, and rewrites the latter in terms of $M_V$ and $\langle v^2 \rangle$. Section~\ref{sec:HEF} discusses the $O(v^2)$ correction to the DLA-HEF resummed coefficient function of Ref.~\cite{Flett:2024htj}. Subsection~\ref{sec:v2-HEF-IF} presents the derivation of the $O(v^2)$ correction to the process-dependent impact factor, while Subsection~\ref{sec:v2-HEF-CF} derives the corresponding resummed correction to the CF coefficient function. Numerical results and the limits of applicability of the obtained expressions are discussed in Sec.~\ref{sec:num}. 
Finally, Sec.~\ref{sec:concl} summarises our conclusions and outlines directions for future work. In the \hyperref[sec:fin-v2-LO]{Appendix A}, we present an expression that resums kinematic higher-order corrections in $v^2$ to the LO coefficient function in $\alpha_s$, providing some insight into the magnitude of relativistic corrections at $O(v^4)$ and beyond. 
In \hyperref[sec:AppB]{Appendix B}, we explain how the $v^2\delta'(x\pm\xi)$ terms in the LO coefficient function in $\alpha_s$ are related to a class of NLL corrections in HEF.

\section{NRQCD matching and different representations for the $O(v^2)$ corrections}\label{sec:NRQCD}

The computation of the $O(v^2)$ correction in NRQCD typically follows the NRQCD matching procedure~\cite{Bodwin:1994jh}. First, one computes the perturbative QCD (pQCD) amplitude for the production of a $Q\bar{Q}$ pair in a particular angular-momentum and spin state, which, for the case of $1^{--}$ quarkonia, is ${}^{2S+1}L^{\text{[col.]}}_J={}^3S^{[1]}_1$, and for a fixed value of the squared relative velocity $v^2$. The resulting $v^2\to 0$ expansion of this amplitude in the rest frame of the $Q\bar{Q}$ pair is then matched to a series of matrix elements of different NRQCD operators with definite velocity scaling:
\begin{eqnarray}
   && {\cal M}^{\text{(pQCD)}}[Q\bar{Q}({}^3S_1^{[1]},\T{ k},\T{ p}=0,S_z)] =  M_{Q\bar{Q}} \label{eq:match-eqn} \\
   && \times\bigg[ O(v^4) + c^{(v^0)}_i \bra{Q\bar{Q}({}^3S_1^{[1]},\T{ k},\T{ p}=0,S_z)} \psi^\dagger \sigma_i \chi \ket{0} \nonumber \\
   && + c^{(v^2)}_i \bra{Q\bar{Q}({}^3S_1^{[1]},\T{ k},\T{ p}=0,S_z)} \psi^\dagger \sigma_i \Big(\frac{-{\mathop{\T{ D}}\limits^{\leftrightarrow}}^2}{m^2_Q} \Big) \chi \ket{0}  \bigg], \nonumber
\end{eqnarray}
where $(\chi)\psi^\dagger$ are the NRQCD creation operators for the heavy (anti-)quark at the point $x=0$, $\sigma_i$ denote the Pauli matrices, and $\mathop{\T{ D}}\limits^{\leftrightarrow}=\big(\mathop{\boldsymbol{ \nabla}}\limits^{\leftarrow} + \mathop{ \boldsymbol{\nabla}}\limits^{\rightarrow}\big)/2 + ig\T{ A}$ is the symmetric QCD covariant derivative. Here, $\T{ k}=(\T{ p}_{Q}-\T{ p}_{\bar{Q}})/2\sim O(v)$ is the relative momentum of the heavy quarks, $\T{ p}=\T{ p}_Q+ \T{ p}_{\bar{Q}}$ is their total momentum, and $M_{Q\bar{Q}}=2\sqrt{m_Q^2 + \T{ k}^2}$. The non-relativistic states on the r.h.s. of eqn.~(\ref{eq:match-eqn}) are normalised as $\braket{Q\bar{Q}({}^3S_1^{[1]},\T{ k},\T{ p},S_z)}{Q\bar{Q}({}^3S_1^{[1]},\T{ k}',\T{ p}',S'_z)}=(2\pi)^{6}\delta^{(3)}(\T{ k}-\T{ k}')\delta^{(3)}(\T{ p}-\T{ p}')\delta_{S_z,S_z'}$, and the overall factor $M_{Q\bar{Q}}$ accounts for the difference in the normalisation of the relativistic plane-wave states on the l.h.s. and the non-relativistic states on the r.h.s.

The short-distance coefficients $c^{(v^0)}$ and $c^{(v^2)}$ can be obtained by enforcing eqn.~(\ref{eq:match-eqn}) order by order in $v^2$ and $\alpha_s$. They are assumed to be independent of the particular state $\ket{Q\bar{Q}({}^3S_1^{[1]},\T{ k},\T{ p}=0,S_z)}$, so that replacing the latter with the physical quarkonium bound state $\ket{V}$ (with $V=J/\psi, \Upsilon,\ldots$) yields the production amplitude for the physical quarkonium:
\begin{eqnarray}
   &&  {\cal A}[V(J_z,\T{ p}=0)] = \sqrt{M_V} \label{eq:match-eqn-phys} \\
   && \times \bigg[ c_i^{(v^0)} \bra{V(J_z,\T{ p}=0)} \psi^\dagger \sigma_i \chi \ket{0}\nonumber \\
   && + c_i^{(v^2)} \bra{V(J_z,\T{ p}=0)} \psi^\dagger \sigma_i \Big(\frac{-{\mathop{\T{ D}}\limits^{\leftrightarrow}}^2}{m^2_Q} \Big) \chi \ket{0} + O(v^4)\bigg],\nonumber
\end{eqnarray}
where the non-relativistic states on the r.h.s. are normalised as $\braket{V(J_z,\T{ p})}{V(J_z',\T{ p}')}=(2\pi)^3 \delta^{(3)}(\T{ p}-\T{ p}')\delta_{J_z,J_z'}$, hence the factor $\sqrt{M_V}$ on the r.h.s. The long-distance matrix elements (LDMEs) $\bra{V} \psi^\dagger \kappa \chi \ket{0}$ should be determined using non-perturbative methods.

It is convenient to introduce the ratio of LDMEs~\footnote{Note that in the case of inclusive production or decays, the matrix elements {\it without} projectors $\ket{0}\bra{0}$ in the middle appear in the definition of $\langle v^2 \rangle^{\text{(incl.)}}$. However, these inclusive LDMEs are related to eqn.~(\ref{eq:v2-def}) through the vacuum-saturation approximation, which is believed to be accurate up to higher-order corrections in $v^2$. }:
\begin{equation}
    \langle v^2 \rangle = -\frac{ \bra{V} \psi^\dagger \sigma_i {\mathop{\T{ D}}\limits^{\leftrightarrow}}^2 \chi \ket{0} \bra{0} \chi^\dagger \sigma_i  \psi \ket{V}}{m^2_Q \left\vert \bra{V} \psi^\dagger \sigma_j \chi \ket{0}\right\vert^2}. \label{eq:v2-def}
\end{equation}

In Ref.~\cite{Gremm:1997dq}, Gremm and Kapustin showed, using the NRQCD equations of motion, that:
\begin{equation}
  \langle v^2 \rangle = \frac{M_{V}-2m_Q^{\text{(pole)}}}{m_Q^{\text{(pole)}}} \big( 1 + O(v^2) \big) , \label{eq:GK-relation}
\end{equation}

where the superscript $^{\text{(pole)}}$ specifies the scheme in which $m_Q$ is defined. However, the scheme ambiguity of $m_Q$ is irrelevant for the computations presented in this paper, because we never encounter UV-divergences related with heavy-quark mass renormalisation. The physical mass $M_V$ is experimentally known, while the parameter $\langle v^2 \rangle$ can be estimated, for example, from potential models~\cite{Bodwin:2006dn, Chung:2020zqc}, with a typical value of $\langle v^2 \rangle\simeq 0.25$ for $J/\psi$, or computed using lattice QCD~\cite{Bodwin:1996tg, Bodwin:2001mk}\footnote{See also a recent fit of $\langle v^2 \rangle$ for $J/\psi$ and $\psi(2S)$ in Ref.~\cite{Bertone:2026lhv}, based on the inclusive hadroproduction data, which gives $\langle v^2 \rangle \simeq 0.22$ for $J/\psi$.}. Therefore, in computations accurate up to $O(v^2)$, it is natural to express all dependence on $m_Q$ in terms of $\langle v^2 \rangle$ and $M_V$ using the relation:
\begin{equation}
    m_Q= \frac{M_V}{2}\Big(1-\frac{\langle v^2 \rangle}{2} + O(v^4) \Big). \label{eq:GK-rel-mQ}
\end{equation}

In the present paper, results expressed solely in terms of $\langle v^2 \rangle$ and $M_V$ will be referred to as being in the {\it physical mass (PM) scheme}, and the corresponding expressions will carry the superscript ${}^{\text{(PM)}}$.

Due to eqn.~(\ref{eq:GK-relation}), one can replace the factor $M_{Q\bar{Q}}$ on the r.h.s. of eqn.~(\ref{eq:match-eqn}) by $M_V$ up to corrections of $O(v^4)$. The advantage of this replacement is that one then needs to expand only the l.h.s. of eqn.~(\ref{eq:match-eqn}) in powers of $v^2$. The resulting expression for the physical amplitude in eqn.~(\ref{eq:match-eqn-phys}) then depends explicitly on $m_Q$, $M_V$, and both LDMEs. This representation of the $O(v^2)$ corrections is widely used in the NRQCD literature. For example, the correction derived in Ref.~\cite{Blask:2025jua} is given in this {\it mixed-mass (MM) scheme}, which will be denoted by the superscript ${}^{\text{(MM)}}$ throughout the present paper. For expressions that are valid in both schemes, the superscript $^{(S)}$ will be used, where $S\in \{\text{MM},\text{PM}\}$.

In practice, to obtain the short-distance coefficients $c^{(v^0,\text{MM})}$ and $c^{(v^2,\text{MM})}$ in the MM scheme, one first computes the pQCD amplitude for the production of a $Q\bar{Q}({}^3S_1^{[1]})$ state, which can be constructed from the usual pQCD amplitude for heavy-quark-pair production with the help of the projector~\cite{Bodwin:2002cfe}:
\begin{eqnarray}
 && v_i(p_{\bar{Q}})\otimes \bar{u}_j(p_Q) \to \frac{\delta_{ij}}{\sqrt{N_c}(E_Q+m_Q)} \nonumber \\
 &&\times \left(\slashed{p}_{\bar{Q}}-m_Q \right) \slashed{\varepsilon}^*(p) \frac{\slashed{p}+2E_Q}{4E_Q} \left( \slashed{p}_{Q} + m_Q \right),\label{eq:NRQCD-proj}
\end{eqnarray}
where $E_Q=\sqrt{m_Q^2+\T{ p}_Q^2}$, $i$ and $j$ are colour indices, and $\varepsilon(p)$ is the spin-polarisation vector. The four-momenta of the quark and antiquark are parametrised as:
\begin{eqnarray}
    p^\mu_Q=\frac{p^\mu}{2}+k^\mu,\;\; p^\mu_{\bar{Q}}=\frac{p^\mu}{2}-k^\mu, \label{eq:def-p-k}
\end{eqnarray}
with $p^2=M_{Q\bar{Q}}^2$ and $k\cdot p=0$. The relative momentum $\T{ k}$ in this parametrisation is canonically conjugate to the separation $\T{ r}$ between the quarks, and the corresponding non-relativistic Hamiltonian is
\begin{equation}
    H_{\text{pot. mod.}}= \frac{\T{ p}^2}{4m_Q} + \frac{\T{ k}^2}{2m_{*}} + V(\T{ r}),
\end{equation}
where $V(\T{ r})$ is the interaction potential and $m_{*}=m_Q/2$ is the effective mass in the two-body problem.

The pQCD amplitude computed using the projector (\ref{eq:NRQCD-proj}) is then expanded in $\T{ k}$ up to second order, and the matching coefficients in eqn.~(\ref{eq:match-eqn-phys}) are obtained via
\begin{eqnarray}
    && \hspace{-6mm}\varepsilon^*_i(S_z)c_i^{(v^0,\text{MM})} = \frac{1}{M_V{\cal N}_{\text{NRQCD}}}  \label{eq:c-v0-gen} \\ && \times {\cal M}^{\text{(pQCD)}}[Q\bar{Q}({}^3S_1^{[1]},\T{ k}=0,\T{ p}=0,S_z)],\nonumber \\
    &&\hspace{-6mm}\varepsilon^*_i(S_z)c_i^{(v^2,\text{MM})} = \frac{\delta_{lj}}{6M_V{\cal N}_{\text{NRQCD}}} \label{eq:c-v2-gen} \\
    && \times \bigg(\frac{\partial}{\partial k_l \partial k_j} {\cal M}^{\text{(pQCD)}}[Q\bar{Q}({}^3S_1^{[1]},\T{ k},\T{ p}=0,S_z)]\bigg)_{\T{ k}=0},\nonumber
\end{eqnarray}
where ${\cal N}_{\text{NRQCD}}^2 = |\bra{Q\bar{Q}({}^3S_1^{[1]},\T{ k}=0,\T{ p}=0,S_z)} \psi^\dagger \sigma_i \chi \ket{0}|^2 = 2N_c(2S+1) = 6N_c$. The factor $1/6$ in eqn.~(\ref{eq:c-v2-gen}) follows from the identity
\[
\int\frac{d\Omega_{\T{ k}}}{4\pi} \frac{k^ik^j}{2} = \frac{\T{ k}^2}{6}\delta_{ij},
\]
where integrating the amplitude over the solid angle $\Omega_{\T{ k}}$ projects out the contribution of the $L=0$ state. The projector $\delta_{ij}$ in eqn.~(\ref{eq:c-v2-gen}) can be expressed covariantly as $-g_{\mu\nu}+p_\mu p_\nu/(4m_Q^2)$.

The leading-order in velocity physical LDME in eqn.~(\ref{eq:match-eqn-phys}) can be determined either from a fully non-perturbative calculation (e.g. using lattice QCD) or from the radial wave function at the origin, $R(0)$, obtained in potential models:
\begin{equation}
    \bra{V(J_z)}\psi^\dagger \sigma_i \chi \ket{0} = {\cal N}_{\text{NRQCD}} \frac{R(0)}{\sqrt{4\pi}} \varepsilon_i^*(J_z)+O(v^2),
\end{equation}
It can also be extracted from the leptonic decay width $\Gamma(V\to l^+l^-)$. In the latter case, the $O(\alpha_s)$ and $O(v^2)$ corrections to the short-distance coefficient of the decay width must be taken into account. The corresponding expression in the {\it PM scheme} (for $N_c=3$) is given in eqn.~(25) of Ref.~\cite{Bodwin:2006yd}:
\begin{eqnarray}
   && \Gamma^{\text{(PM)}}(V\to l^+l^-) = \frac{8\pi\alpha^2 e_Q^2}{9} \frac{ |\bra{V}\psi^\dagger \sigma_i \chi \ket{0}|^2}{M_V^2} \nonumber \\
   && \times\Big[1 - \frac{8}{3\pi}\alpha_s(\mu_R) - \frac{\langle v^2 \rangle}{6} \Big]^2.
\end{eqnarray}

\section{The $O(v^2)$ correction to the LO coefficient function in $\alpha_s$} \label{sec:LO-CF-v2-corr}

For the quarkonium photoproduction process
\begin{equation}
    \gamma(q) + h(P) \to V(p) + h(P'), \label{eq:proc}
\end{equation}
where the target hadron $h$ may be either a proton or a nucleus, the present paper closely follows the notation of Ref.~\cite{Flett:2024htj}, in which the photoproduction amplitude is written as
\begin{eqnarray}
 && \mathcal A^{(S)}=-\varepsilon_{\perp j}^{\gamma}\bra{V(J_z)}\psi^\dagger \sigma_j \chi \ket{0} \label{eq:Ampl-CF} \\
 && \times  \sum\limits_{i={q,g}} \int\limits_{-1}^1\!\! \frac{dx}{x^{1+\delta_{i,g}}}  C_{i}^{(S)}\left(x,\xi;\mu_F, \mu_R\right) {F}_{i}(x,\xi,t;\mu_F),\! \nonumber
\end{eqnarray}
where $F_i(x,\xi,t,\mu_F)$ denotes the GPD of the target hadron $h$, $t=(P'-P)^2$ is the momentum transfer, assumed to satisfy $-t\ll M_V^2$ within the region of applicability of CF, the skewness parameter is $\xi=q\cdot(P-P')/(q\cdot(P+P'))\simeq M_V^2/(2s)$, $s=(P+q)^2$, and $C_i(x,\xi,\mu_F,\mu_R)$ is the CF coefficient function.

The well-known LO coefficient function in $\alpha_s$ for the process (\ref{eq:proc}) is
\begin{equation}
    C_g^{(0,S)}(x,\xi)=\frac{x^2 c^{(S)}}{(x+\xi-i\varepsilon) (x-\xi+i\varepsilon)}.\label{eq:Cg-LO}
\end{equation}
The overall factor depends on the representation $(S)$ adopted for the $v^2$ correction. In the MM scheme it is given by
\begin{equation}
    c^{\text{(MM)}}=\frac{(4\pi)^{3/2} \alpha_s(\mu_R) e e_q}{m_Q \sqrt{N_c M_V}{\cal N}_{\text{NRQCD}}}, \label{eq:cMM-def}
\end{equation}
while in the PM scheme it can be written as
\begin{equation}
    c^{\text{(PM)}}=\frac{2(4\pi)^{3/2} \alpha_s(\mu_R) e e_q}{M_V^{3/2} \sqrt{N_c}{\cal N}_{\text{NRQCD}}}. \label{eq:cGK-def}
\end{equation}

The imaginary part of the LO coefficient function (\ref{eq:Cg-LO}) is
\begin{equation}
    \text{Im}\; C_g^{(0,S)}(x,\xi) =  -\frac{\pi x^2 c^{(S)}}{2\xi} \bigg[ \delta(x-\xi) + \delta(x+\xi) \bigg].
\end{equation}

In the same notation, the $O(v^2)$ correction to the LO coefficient function is
\begin{equation}
  C_g^{(v^2,S)}(x,\xi)= \langle v^2 \rangle \frac{x^2 c^{(S)} g^{(S)}(x,\xi) }{(x-\xi+i\varepsilon)^2 (x+\xi-i\varepsilon)^2}, \label{eq:v2-C-LO-exact}
\end{equation}
where the function $g^{(S)}(x,\xi)$ corresponding to the MM-scheme result was computed in Ref.~\cite{Blask:2025jua} and cross-checked by the author of the present paper (see the \hyperref[sec:fin-v2-LO]{Appendix A}):
\begin{equation}
    g^{\text{(MM)}}(x,\xi)=\frac{7\xi^2 - x^2}{3}.\label{eq:gMM-LO}
\end{equation}
To obtain the PM-scheme result, one simply expresses the $m_Q$ dependence in eqn.~(\ref{eq:cMM-def}) using eqn.~(\ref{eq:GK-rel-mQ}) and adds the resulting $O(v^2)$ terms to the MM-scheme result given by eqns.~(\ref{eq:v2-C-LO-exact}) and (\ref{eq:gMM-LO}). The corresponding function $g^{(S)}(x,\xi)$ is
\begin{equation}
     g^{\text{(PM)}}(x,\xi)=\frac{11\xi^2 + x^2}{6}. \label{eq:gPM-LO}
\end{equation}

The presence of second-order poles in eqn.~(\ref{eq:v2-C-LO-exact}) requires the GPD $F_g(x,\xi,t,\mu_F)$ to be a smooth function at the points $x=\pm\xi$. This property may not be preserved under GPD evolution and may indicate a mild violation of collinear factorisation for the real part of the amplitude, as already noted in Ref.~\cite{Blask:2025jua}. The imaginary part of the amplitude is more robust with respect to this issue, since it does not develop a divergent contribution even if the GPD is not smooth at $x=\pm\xi$. The imaginary part of the coefficient function (\ref{eq:v2-C-LO-exact}) is
\begin{eqnarray}
   && \text{Im}\; C_g^{(v^2,S)}(x,\xi) = -\langle v^2\rangle\frac{\pi x^2 c^{(S)}}{2\xi} \nonumber \\
    && \times \bigg[ b_0^{(S)}\big(\delta(x-\xi)+\delta(x+\xi)\big) \nonumber \\
    && - \big(\delta'(x-\xi)-\delta'(x+\xi) \big)\frac{g^{(S)}(\xi,\xi)}{{2}\xi} \bigg],\label{eq:Cg-LO-Im}
\end{eqnarray}
where the coefficient
\begin{eqnarray}
    b_0^{(S)} &=& -\frac{1}{{2}\xi}\bigg[ \frac{g^{(S)}(x,\xi)}{\xi} - \frac{\partial g^{(S)}(x,\xi)}{\partial x} \bigg]_{x=\xi} \nonumber \\
    &=& \left\{\begin{array}{cc}
        -4/3 & \text{for }S=\text{MM},  \\
        -5/6 & \text{for }S=\text{PM}.
    \end{array} \right. \label{eq:b0-coeffs}
\end{eqnarray}


The terms proportional to $\delta'(x\pm\xi)$ in eqn.~(\ref{eq:Cg-LO-Im}) are not power-suppressed in the high-energy limit $\xi\ll 1$, despite the presence of an additional factor of $\xi$. Indeed, for $F_g(x,\xi)\sim x^{-\lambda}$, one has
$\xi \left.\frac{\partial_x F_g(x,\xi)}{F_g(x,\xi)}\right|_{x=\xi}=-\lambda$.
However, in phenomenological applications one often finds $\lambda\lesssim 1$ at the scale $\mu_F\sim M_V$, so the numerical contribution of the $\delta'(x\pm\xi)$ terms is generally subleading. In Appendix~\ref{sec:AppB}, we demonstrate that the HEF-resummed corrections associated with these terms are of NLL order ($\propto \alpha_s^{n+1}\ln^{n-1}(x/\xi)$) and therefore lie beyond the logarithmic accuracy considered in the present paper.

One also notes that the coefficients in eqn.~(\ref{eq:b0-coeffs}) are negative. The fact that the $O(v^2)$ correction decreases the imaginary part of the amplitude is well known. At small $\xi$, this behaviour has also been observed in the CGC calculation of Ref.~\cite{Mantysaari:2022kdm}. Within collinear factorisation, the same qualitative behaviour has been reported for deeply virtual electroproduction of quarkonia~\cite{Hoodbhoy:1996zg}, as well as for photoproduction in a formalism employing $\T{k}_T$-dependent light-cone distribution amplitudes for the vector meson~\cite{Goloskokov:2005sd}.

\section{The $O(v^2)$ correction to the HEF-resummed coefficient function}\label{sec:HEF}

The computation of the $O(v^2)$-correction to the HEF-resummed coefficient function of Ref.~\cite{Flett:2024htj} consists of two steps. First, the $O(v^2)$-correction to the process-dependent impact-factor (or HEF coefficient function -- $h^{(v^2)}(\T{ q}_T^2)$) is computed in Sec.~\ref{sec:v2-HEF-IF}. Then in Sec.~\ref{sec:v2-HEF-CF} this process-dependent factor is convoluted with the universal HEF resummation function ${\cal C}_{ig}(\xi/x,\T{ q}_T^2,\mu_F^2)$ to obtain the result for the $O(v^2)$-correction to the DLA-HEF-resummed CF coefficient function: $C_{i}^{(\text{HEF},v^2,S)}(x,\xi)$.

\subsection{The $O(v^2)$ correction to the impact factor\label{sec:v2-HEF-IF}}

\begin{figure}
    \centering
    \includegraphics[width=0.9\linewidth]{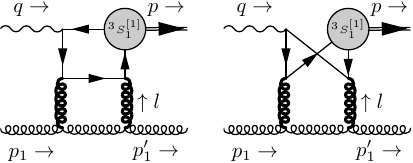}
    \caption{Typical one-loop Feynman diagrams contributing to the Regge limit of the partonic process (\ref{proc:ga+g-cc+g}) at NLO. All 6  diagrams with both $t$-channel gluons attached to the quark line should be included. The thick $t$-channel gluons are Glaubers.}
    \label{fig:diags}
\end{figure}

In the computation of the HEF impact factor at $O(v^2)$, the present paper follows the same strategy as that used for the computation of the $O(v^0)$ impact factor, described in the appendix of Ref.~\cite{Flett:2024htj}. One considers the pQCD amplitudes for the NLO CF subprocesses
\begin{eqnarray}
    && \gamma(q)+g(p_1)\to Q\bar{Q}\left[{}^3S_1^{[1]} \right](p,k) + g(p_1'), \label{proc:ga+g-cc+g} \\
    && \gamma(q)+q(p_1)\to Q\bar{Q}\left[{}^3S_1^{[1]} \right](p,k) + q(p_1'), \label{proc:ga+q-cc+q}
\end{eqnarray}
at one loop and in the Regge limit, $\hat{s}=(q+p_1)^2\gg m_Q^2$, $\hat{t}=(p_1-p_1')^2=0$, and $\xi\ll x=(q\cdot p_1)/(q\cdot P)$. The leading-power contribution in this limit is given by the one-loop box diagrams shown in Fig.~\ref{fig:diags}, with the heavy-quark and antiquark spinors replaced by the projector (\ref{eq:NRQCD-proj}), represented in the diagrams by a grey blob. The leading contribution arises from the Glauber region of integration over the $t$-channel gluon momentum $l$, where one has $|l_+l_-|\ll \T{ l}_T^2$. (We define the Sudakov components of a vector $k^\mu$ as $k_{\pm}=n_{\pm}\cdot k$, where $n_+^\mu=2q^\mu/\sqrt{s}$ and $n_-^\mu=2P^\mu/\sqrt{s}$, $n_{\pm}\cdot k_T=0$, and the upper/lower position of the $\pm$ indices carries no meaning in the notation of the present paper.) In this region, one may replace the Feynman-gauge $t$-channel gluon propagators using Gribov's prescription:
\begin{equation}
   \frac{ -ig^{\mu\nu}}{l^2+i\varepsilon} \to \frac{i}{2 \T{ l}_T^2}(n_-^{\mu} n_+^{\nu}+n_+^{\mu} n_-^{\nu}),\label{eq:Gribov}
\end{equation}
after which the leading-power contribution to the one-loop amplitudes for the processes (\ref{proc:ga+g-cc+g}) and (\ref{proc:ga+q-cc+q}) factorises as
\begin{eqnarray}
    && {\cal M}^{\text{(pQCD)}}_{\gamma i} = \int \frac{i\, d^{2-2\epsilon}\T{ l}_T}{4(2\pi)^{4-2\epsilon}} \int\limits_{-\infty}^{+\infty} dl_+ \frac{{\cal A}_{ab}(l_+,\T{ l}_T,k)}{2\T{ l}_T^2} \nonumber \\
    && \times \int\limits_{-\infty}^{+\infty}dl_- \frac{{\cal B}^{ab}_{i}(l_-,\T{ l}_T)}{2\T{ l}_T^2},\label{eq:ampl_ga+i-cc+i}
\end{eqnarray}
with $i=g$ for the subprocess (\ref{proc:ga+g-cc+g}) and $i=q$ for (\ref{proc:ga+q-cc+q}). The partonic target impact factors ${\cal B}^{ab}_{i}$ are discussed in detail in the appendix of Ref.~\cite{Flett:2024htj}. The task of the present paper is to compute the $O(\T{ k}^2)\sim O(v^2)$ contribution to the projectile impact factor ${\cal A}_{ab}$.

The Sudakov decomposition of the photon and $Q\bar{Q}$-pair momenta is
\begin{eqnarray}
    && q^\mu = \frac{q_-}{2} n_+^\mu, \label{eq:q-photon} \\
    && p^\mu = \frac{1}{2}\left( q_- n_+^\mu + \frac{M_{Q\bar{Q}}^2}{q_-} n_-^\mu \right),\label{eq:p-pair}
\end{eqnarray}
where $q_-=\sqrt{s}$.
 
The velocity expansion of the impact factor can be defined as follows:
\begin{eqnarray}
  &&  \int\frac{d\Omega_{\T{ k}}}{4\pi} {\cal A}_{ab}(l_+,\T{ l}_T,k) = {\cal A}_{ab}^{(v^0,S)}(l_+,\T{ l}_T) \nonumber \\
  && + v^2 {\cal A}_{ab}^{(v^2,S)}(l_+,\T{ l}_T)+O(v^4),
\end{eqnarray}
with $v^2=\T{ k}^2/m_Q^2$, where
\begin{eqnarray}
     &&{\cal A}^{(v^0,\text{MM})}_{ab}(l_+,\T{ l}_T) =  \frac{1}{\sqrt{M_V}}{\cal A}_{ab}(l_+,\T{ l}_T,k=0), \\
     && {\cal A}^{(v^2,\text{MM})}_{ab}(l_+,\T{ l}_T) = \frac{m_Q^2}{6\sqrt{M_V}} \left(-g_{\mu\nu} + \frac{p_\mu p_\nu}{4m_Q^2} \right) \nonumber \\
     && \times\left[\frac{\partial^2}{\partial k_\mu \partial k_\nu} {\cal A}_{ab}(l_+,\T{ l}_T,k) \right]_{ k=0}.
\end{eqnarray}

The diagrammatic calculation, taking into account the projector (\ref{eq:NRQCD-proj}), gives
\begin{eqnarray}
  &&\hspace{-6mm}{\cal A}^{(v^0,\text{MM})}_{ab}(l_+,\T{ l}_T) = \frac{32c^{\text{(MM)}} q_-^2\T{ l}_T^2 \delta_{ab} g^{\mu\nu}_\perp \varepsilon_\mu^*(p) \varepsilon_\nu(q)}{ D_1 D_2}, \\
&&\hspace{-6mm}{\cal A}^{(v^2,\text{MM})}_{ab}(l_+,\T{ l}_T) = -\frac{512c^{\text{(MM)}} q_-^2 \T{ l}_T^2 \delta_{ab} g^{\mu\nu}_\perp \varepsilon_\mu^*(p) \varepsilon_\nu(q) }{3 D_1^3 D_2^3} \nonumber \\
&&\times \Big[ 16(\T{ l}_T^2 + m_Q^2)^2 (2\T{ l}_T^2+4m_Q^2\T{ l}_T^2+13m_Q^4) \nonumber \\
&& - 4 l_+^2 q_-^2 (7\T{ l}_T^2+22m_Q^2 \T{ l}_T^2 + 14m_Q^4) + l_+^4q_-^4 \Big].
\end{eqnarray}
where
$
D_1=-i\varepsilon-2l_+q_-+4(m_Q^2+\T{ l}_T^2)$, 
$D_2=-i\varepsilon+2l_+q_-+4(m_Q^2+\T{ l}_T^2)$,
and
$
g^{\mu\nu}_\perp=g^{\mu \nu}- \frac{n_-^{\nu} n_+^{\mu}+n_+^{\nu} n_-^{\mu}}{2}$.

In both cases, the $l_+$ integral can be evaluated by taking the residue at
\begin{equation}
l^+_{*} = \frac{-4(m_Q^2+\T{ l}_T^2) + i\varepsilon}{2q_-},\label{eq:lplus-star-v0}
\end{equation}
so that the velocity expansion of the $l_+$-integrated impact factor can be written as
\begin{eqnarray}
    && \int\limits_{-\infty}^{+\infty} dl_+  \int\frac{d\Omega_{\T{ k}}}{4\pi} {\cal A}_{ab}(l_+,\T{ l}_T,k) \nonumber \\
    && = 4\pi i c^{(\text{MM})} \delta_{ab} (\varepsilon^{*}_\mu(p) \varepsilon_\nu (q) g^{\mu\nu}_\perp) \nonumber \\
    && \hspace{-5mm}\times \frac{q_- {\T{ l}_T^2}}{m_Q^2} \Big( h^{(v^0,\text{MM})}(\T{ l}_T^2) + v^2 h^{(v^2)}(\T{ l}_T^2) +O(v^4)  \Big), \label{eq:int-A-IF}
\end{eqnarray}
where:
\begin{eqnarray}
  && \hspace{-5mm} h^{(v^0,\text{MM})}(\T{ q}_T^2) = \frac{m_Q^2}{m_Q^2+ \T{ q}_T^2},\label{eq:h-v0-MM} \\
&& \hspace{-5mm} h^{(v^2)}(\T{ q}_T^2) = -\frac{m_Q^2 \left(7 m_Q^4+5 m_Q^2 \T{ q}_T^2+ 2 (\T{ q}_T^2)^2\right)}{3 \big(m_Q^2+\T{ q}_T^2\big)^3}. \label{eq:h-v2-00}
\end{eqnarray}
Equation~(\ref{eq:h-v0-MM}) coincides with the $O(v^0)$ HEF coefficient function, eqn.~(13) of Ref.~\cite{Flett:2024htj}. Equation~(\ref{eq:h-v2-00}) is not yet the final MM-scheme result for the coefficient of the $O(v^2)$ correction to the impact factor. One must still take into account the fact that the overall factor $q_-$ in eqn.~(\ref{eq:int-A-IF}), together with the factor $P_+$ from the target impact factor ${\cal B}$, forms the overall factor of the squared centre-of-mass energy in the amplitude (\ref{eq:ampl_ga+i-cc+i}), which, for a fixed value of $\xi$, also depends on $M_{Q\bar{Q}}$ and hence on $v^2$:
\begin{equation}
  s= q_- P_+ = \frac{M_{Q\bar{Q}}^2}{2\xi}  = \frac{2m_Q^2}{\xi} (1+v^2) + O(v^4).
\end{equation}

Including this $O(v^2)$ contribution in the total relativistic correction, one finally obtains the impact factor in the MM scheme:
\begin{eqnarray}
   h^{(v^2,\text{MM})}(\T{ q}_T^2) = -\frac{m_Q^2(4m_Q^4-m_Q^2\T{ q}_T^2-(\T{ q}_T^2)^2)}{3(m_Q^2+\T{ q}_T^2)^3}.
\end{eqnarray}

To rewrite the result in the PM scheme, one simply re-expresses all $m_Q$ dependence in eqn.~(\ref{eq:int-A-IF}) using eqn.~(\ref{eq:GK-rel-mQ}) and collects all terms of $O(v^2)$ together. Thus-obtained coefficients of the velocity expansion of the impact factor in the PM scheme are
\begin{eqnarray}
  && \hspace{-10mm}h^{(v^0,\text{PM})}(\T{ q}_T^2) = \frac{M_V^2}{M_V^2+4\T{ q}_T^2}, \label{eq:h-v0-PM} \\
  &&\hspace{-10mm}h^{(v^2,\text{PM})}(\T{ q}_T^2) = -\frac{M_V^2\big(5M_V^4-8M_V^2\T{ q}_T^2+16(\T{ q}_T^2)^2\big)}{6(M_V^2+4\T{ q}_T^2)^3}. \label{eq:h-v2-PM}
\end{eqnarray}

As will be explained below, an important observation is that, in both schemes, the $\T{ q}_T^2\to 0$ limit of the impact factor reproduces the corresponding coefficients in eqn.~(\ref{eq:b0-coeffs}):
\begin{equation}
    h^{(v^2,S)}(0)=b_0^{(S)}.\label{eq:b0-h-eqn}
\end{equation}

\begin{figure}
    \centering
    \includegraphics[width=0.95\linewidth]{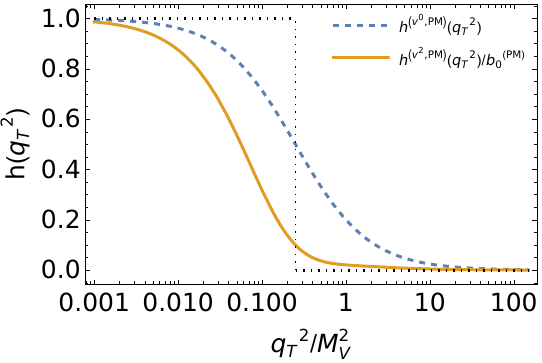}
    \caption{Plots of the functions $h^{(v^0,\text{PM})}(\T{q}_T^2)$ (dashed line) and $h^{(v^2,\text{PM})}(\T{q}_T^2)/b_0^{(\text{PM})}$ (solid line) as functions of $\T{q}_T^2/M_V^2$. The function $\theta(\T{q}_T^2<M_V^2/4)$ appearing in eqn.~(\ref{eq:b1-h-expr}) is shown by the dotted line for comparison.}
    \label{fig:h-plots}
\end{figure}

The HEF coefficient functions (\ref{eq:h-v0-PM}) and (\ref{eq:h-v2-PM}) are plotted in Fig.~\ref{fig:h-plots}. One can see that the function (\ref{eq:h-v2-PM}) is concentrated primarily in the region $\T{q}_T^2\lesssim 0.1M_V^2$, which turns out to be important for the behaviour of the $O(v^2)$ correction to the resummed CF coefficient function, as discussed in the next subsection.

\subsection{The $O(v^2)$ HEF-resummed CF coefficient function}\label{sec:v2-HEF-CF}

As explained in Refs.~\cite{Flett:2024htj,Ivanov:2007je}, the HEF resummation contributes to the imaginary part of the CF coefficient function in the DGLAP region at leading power in $\rho=\xi/|x|\ll 1$. The corresponding factorisation formula, including the $O(v^2)$ correction and valid in the LLA ($\alpha_s^{n}\ln^{n-1}(1/\rho)/\rho$) and next-to-LLA ($\alpha_s^{n+1}\ln^{n-1}(1/\rho)/\rho$), takes the form
\begin{eqnarray}
   && C^{(\text{HEF},S)}_{i}(\rho)= \frac{-i\pi}{2} \frac{c^{(S)}}{\rho} \int\limits_0^{\infty} d\T{ q}_T^2\ {\cal C}_{gi}(\rho,\T{ q}_T^2,\mu_F^2) \nonumber \\
   &&\times \Big( h^{(v^0,S)}(\T{ q}_T^2) + \langle v^2 \rangle h^{(v^2,S)}(\T{ q}_T^2) \Big).\label{eq:Master-formula-HEF}
\end{eqnarray}

It is convenient to introduce the Mellin transform with respect to $\rho$:
\begin{equation}
    C(\rho) = \int\frac{dN}{2\pi i} \rho^{-N-1} C(N),\label{eq:C-rho-Mellin}
\end{equation}
which maps poles $1/N^n$ in the Mellin plane to terms proportional to $\ln^{n-1}(1/\rho)/[(n-1)!\rho]$. The DLA-HEF resummation factor (see, e.g., Ref.~\cite{Lansberg:2021vie} and the references therein for details), which appears in eqn.~(\ref{eq:Master-formula-HEF}), has a particularly simple form in this representation:
\begin{equation}
    {\cal C}_{gg}(N,\T{ q}_T^2) = \frac{\gamma_N}{\T{ q}_T^2} \left(\frac{\T{ q}_T^2}{\mu_F^2} \right)^{\gamma_N}, \label{eq:res-fact-Mellin}
\end{equation}
with $\gamma_N=\hat{\alpha}_s(\mu_R)/N$, $\hat{\alpha}_s(\mu_R)=\alpha_s(\mu_R) C_A/\pi$. In the quark-induced channel, the LLA resummation function is
\begin{equation}
    {\cal C}_{gq}(N,\T{q}_T^2,\mu_F^2)=\frac{2C_F}{C_A}\left[  {\cal C}_{gg}(N,\T{q}_T^2,\mu_F^2)
- \delta(\T{q}_T^2)\right].  \label{eq:C-gq}
\end{equation}

The velocity expansion of the Mellin-space resummed CF coefficient function takes the form
\begin{equation}
    C_i^{(\text{HEF},S)}(N) = C_i^{(\text{HEF},v^0,S)}(N) + \langle v^2 \rangle C_i^{(\text{HEF},v^2,S)}(N),
\end{equation}
where the coefficients of the expansion {in the gluon channel} are
\begin{eqnarray}
 &&\hspace{-7mm} C^{\text{(HEF, $v^{0}$, S)}}_{g}(N) = \frac{-i\pi c^{(S)}}{2} \left(\frac{M_V^2}{4\mu_F^2}\right)^{\gamma_N} \frac{\pi\gamma_N}{\sin (\pi \gamma_N)}, \label{eq:CHEF-v0-Mellin} \\
 &&\hspace{-7mm} C^{\text{(HEF, $v^{2}$,S)}}_{g}(N)= \left(b_0^{(S)}+b_1^{(S)}\gamma_N+b^{(S)}_2 \gamma^2_N\right) \nonumber \\
  && \times C^{\text{(HEF, $v^{0}$, S)}}_{g}(N). \label{eq:CHEF-v2-Mellin}
\end{eqnarray}
The Mellin-space coefficient functions in the quark channel are obtained with the help of eqn.~(\ref{eq:C-gq}) as:
\begin{eqnarray}
 &&   C^{\text{(HEF, $v^{0}$, S)}}_{q}(N) \nonumber \\
 && = \frac{2C_F}{C_A} \Big[ C^{\text{(HEF, $v^{0}$, S)}}_{g}(N) + \frac{i\pi c^{\text{(S)}}}{2}  \Big], \\
 &&   C^{\text{(HEF, $v^{2}$, S)}}_{q}(N) \nonumber \\
 && = \frac{2C_F}{C_A} \Big[ C^{\text{(HEF, $v^{0}$, S)}}_{g}(N) + \frac{i\pi c^{\text{(S)}}}{2} b_0^{\text{(S)}} \Big].
\end{eqnarray}
The coefficients $b_0^{(S)}$ are given in eqn.~(\ref{eq:b0-coeffs}), while the coefficients $b_1^{\text{(MM)}}$ and $b_2^{\text{(MM)}}$ are
\begin{eqnarray}
  b_1^{\text{(MM)}}=\frac{7}{3},\;  b_2^{\text{(MM)}}=-\frac{2}{3},
\end{eqnarray}
whereas in the PM scheme these coefficients are
\begin{eqnarray}
b_1^{\text{(PM)}}=\frac{4}{3},\; b_2^{\text{(PM)}}=-\frac{2}{3}.
\end{eqnarray}

Expanding the Mellin-space resummed coefficient functions (\ref{eq:CHEF-v0-Mellin}) and (\ref{eq:CHEF-v2-Mellin}) in $\alpha_s$ and transforming the result back to $\rho$ space, one obtains
\begin{eqnarray}
   && \hspace{-5mm}\frac{2C^{\text{(HEF, $v^{0}$,S)}}_{g}(\rho) }{-i\pi c^{(S)}} = \delta(\rho-1) + \frac{\hat{\alpha}_s}{\rho} L_\mu  \label{eq:Crho-v0-exp-as}  \\
   && + \frac{\hat{\alpha}^2_s}{\rho} \ln \frac{1}{\rho} \left( \frac{L_\mu^2}{2} + \frac{\pi^2}{6} \right) + O(\alpha_s^3), \nonumber \\
   && \hspace{-5mm}\frac{2C^{\text{(HEF, $v^{2}$, S)}}_{g}(\rho) }{-i\pi c^{(S)}}= b^{(S)}_0\delta(\rho-1) + \frac{\hat{\alpha}_s}{\rho} \left(  b_0^{(S)}L_\mu +b_1^{(S)} \right) \label{eq:Crho-v2-exp-as} \\
   && + \frac{\hat{\alpha}^2_s}{\rho} \ln \frac{1}{\rho} \left( \frac{b_0^{(S)}}{2} L_\mu^2 + b_1^{(S)}L_\mu + b_2^{(S)} + b_0^{(S)}\frac{\pi^2}{6} \right) + O(\alpha_s^3). \nonumber
\end{eqnarray}
The expansion (\ref{eq:Crho-v0-exp-as}) first appeared in Ref.~\cite{Flett:2024htj}. In eqn.~(\ref{eq:Crho-v2-exp-as}), the coefficient of the LO term proportional to $\delta(\rho-1)$ is $b_0^{(S)}$, i.e. the same coefficient that appears in the imaginary part of the LO coefficient function (\ref{eq:Cg-LO-Im}), as expected. Since the DLA HEF coincides with the complete LLA of HEF up to NNLO in $\alpha_s$~\cite{Lansberg:2021vie}, all terms written explicitly in eqns.~(\ref{eq:Crho-v0-exp-as}) and (\ref{eq:Crho-v2-exp-as}) constitute exact predictions for the corresponding terms in the $\alpha_s$ and velocity expansions of the CF coefficient function.

As noted in eqn.~(\ref{eq:b0-h-eqn}), the coefficient $b_0^{(S)}$ is simply given by the value of $h^{(v^2,S)}(\T{q}_T^2)$ at $\T{q}_T^2\to 0$. The coefficients $b_1^{(S)}$ and $b_2^{(S)}$ can likewise be related directly to the $\T{q}_T$ dependence of the impact factor:
\begin{eqnarray}
 &&\hspace{-10mm} b_1^{(S)}=\int\limits_0^{\infty}\frac{d\T{q}^2_T}{\T{q}^2_T} \Big[ h^{(v^2,S)}(\T{q}_T^2) - b_0^{(S)} \theta(\T{q}_T^2<M_V^2/4) \Big], \label{eq:b1-h-expr} \\
 &&\hspace{-10mm} b_2^{(S)}=\int\limits_0^{\infty}\frac{d\T{q}^2_T}{\T{q}^2_T} \ln\bigg[\frac{4\T{q}^2_T}{M_V^2}\bigg] \Big[ h^{(v^2,S)}(\T{q}_T^2) - \frac{b_0^{(S)}M_V^2}{M_V^2+4\T{q}_T^2}  \Big]. \label{eq:b2-h-expr}
\end{eqnarray}
These expressions, together with the plots in Fig.~\ref{fig:h-plots}, help to explain the signs of the coefficients $b_1^{(\text{PM})}$ and $b_2^{(\text{PM})}$ as a consequence of the fact that the function $h^{(v^2,\text{PM})}(\T{q}_T^2)$ is concentrated predominantly in the region of low transverse momenta, $\T{q}_T^2\lesssim 0.1 M_V^2$. For example, if it happened that $h^{(v^2,S)}(\T{q}_T^2)\propto h^{(v^0,S)}(\T{q}_T^2)$, then both integrals (\ref{eq:b1-h-expr}) and (\ref{eq:b2-h-expr}) would vanish, implying $b_1^{(S)}=b_2^{(S)}=0$. This example illustrates that the effect of relativistic corrections is more subtle than a simple overall multiplicative factor in the amplitude.

The inverse Mellin transform of the resummed coefficient function (\ref{eq:CHEF-v2-Mellin}) can be expressed in terms of derivatives of the function $C^{\text{(HEF, $v^{0}$,S)}}_{g}(\rho, \mu_F^2)$ with respect to the factorisation scale:
\begin{eqnarray}
  && C^{\text{(HEF, $v^{2}$,S)}}_{g}(\rho, \mu_F^2) =   \Big[ b_0^{(S)} - b_1^{(S)} \frac{\partial}{\partial\ln\mu_F^2}  \nonumber \\
  &&+ b_2^{(S)} \frac{\partial^2}{\partial(\ln\mu_F^2)^2}  \Big] C^{\text{(HEF, $v^{0}$,S)}}_{g}(\rho, \mu_F^2).  \label{eq:CHEF-deriv-form}
\end{eqnarray}
This representation is useful because the following closed-form expression for the $O(v^0)$ resummed coefficient function in $\rho$ space was derived in Ref.~\cite{Flett:2024htj}:
\begin{eqnarray}
   && \Check{C}_i^{\text{(HEF, $v^0$, S)}}(\rho,\mu_F) = \frac{-i\pi c^{(S)} \hat{\alpha}_s}{2\rho} \bigg\{ A(L_\rho, L_\mu) \nonumber \\
    && \hspace{-5mm}- 2 \sum\limits_{k=1}^\infty \text{Li}_{2k}(-1) B_k(L_\rho,L_\mu) \bigg\}\Big(\delta_{i,g} + \frac{2C_F}{C_A} \delta_{i,q} \Big),\label{eq:CHEF-v0-rho-closed-form}
\end{eqnarray}
with $L_\rho=\hat{\alpha}_s \ln(1/\rho)$ and
\begin{eqnarray}
    A&=&\sqrt{\frac{L_\mu}{L_\rho}} I_1 \big(2\sqrt{L_\mu L_\rho} \big), \\
    B_k&=& \left( \frac{L_\rho}{L_\mu}\right)^{k} \sqrt{\frac{L_\mu}{L_\rho}} I_{2k-1}\big( 2 \sqrt{L_\mu L_\rho} \big),
\end{eqnarray}
where $I_n(x)$ are modified Bessel functions, and the check in the notation $\Check{C}_i^{\text{(HEF, $v^0$, S)}}$ indicates that this expression does not include the LO term in $\alpha_s$, proportional to $\delta(1-\rho)$, thereby avoiding double counting with the LO CF contribution. Consequently, the $O(v^2)$ correction to the resummed coefficient function can be expressed in terms of derivatives of the same closed-form expression by means of eqn.~(\ref{eq:CHEF-deriv-form}).

A useful way to understand the relative magnitude of the $O(v^2)$ correction is to consider its $\rho\ll 1$ asymptotic behaviour, which is determined by the rightmost pole of $C(N)$ of eqn.~(\ref{eq:C-rho-Mellin}) in the complex $N$ plane. The poles of the resummed coefficient functions (\ref{eq:CHEF-v0-Mellin}) and (\ref{eq:CHEF-v2-Mellin}) are located at ($k$ is an integer)
\begin{equation}
 N_k=\frac{\hat{\alpha}_s}{k},
\end{equation}
so that the rightmost pole is located at $N_1=\hat{\alpha}_s$. The residue at this pole gives
\begin{eqnarray}
 &&\hspace{-7mm} \rho\ll 1 :\;   C^{\text{(HEF, $v^{0}$, S)}}_{i}\simeq \frac{-i\pi c^{(S)}}{2\rho} \frac{\hat{\alpha}_s M_V^2}{4\mu_F^2} \rho^{-\hat{\alpha}_s}, \label{eq:CHEF-v0-rho-asy} \\
  &&\hspace{-7mm}  \rho\ll 1 :\; C^{\text{(HEF, $v^{2}$, S)}}_{i}\simeq\big(b_0^{(S)}+b_1^{(S)}+b_2^{(S)} \big)\nonumber \\
  &&\times C^{\text{(HEF, $v^{0}$, S)}}_{i}(\rho,\mu_F). \label{eq:CHEF-v2-rho-asy}
\end{eqnarray}
One might be surprised by the factor $1/\mu_F^2$ in eqn.~(\ref{eq:CHEF-v0-rho-asy}), which arises as an artefact of taking the $\rho\ll 1$ asymptotic limit of the resummed expressions (\ref{eq:CHEF-v0-Mellin}) and (\ref{eq:CHEF-v0-rho-closed-form}). However, one can show that the same $\mu_F$ dependence follows from the evolution of gluon GPDs in the asymptotic $\xi\ll |x|\ll 1$ limit in the fixed-coupling approximation and reflects the very strong scale dependence of GPDs in this regime.

In the PM scheme, the factor appearing in eqn.~(\ref{eq:CHEF-v2-rho-asy}) is
\begin{equation}
b_0^{(\text{PM})}+b_1^{(\text{PM})}+b_2^{(\text{PM})}=-\frac{1}{6},
\end{equation}
so that the resummed coefficient function including the $O(v^2)$ correction behaves as
\begin{eqnarray}
    \rho\ll 1:\;  C^{(\text{HEF, PM})}_{i}(\rho,\mu_F) &\simeq& \Big(1-\frac{\langle v^2 \rangle}{6} \Big) \label{eq:CHEF-PM-v2-asy} \\
    &&\times C^{(\text{HEF}, v^0, \text{PM})}_{i}(\rho,\mu_F). \nonumber
\end{eqnarray}
Hence, the $O(v^2)$ correction to the resummed coefficient function is five times smaller than the corresponding correction to the LO coefficient function (\ref{eq:Cg-LO-Im}), which is governed solely by the coefficient $b_0^{(\text{PM})}$. This is a non-trivial dynamical effect, which can be traced back, with the aid of eqns.~(\ref{eq:b1-h-expr}) and (\ref{eq:b2-h-expr}), to the $\T{q}_T$ dependence of the impact factor $h^{(v^2,\text{PM})}(\T{q}_T^2)$ shown in Fig.~\ref{fig:h-plots}. The conclusions drawn here are further supported by the numerical study of the resummed coefficient function presented in the next subsection.

\subsection{Numerical results}\label{sec:num}

In Ref.~\cite{Flett:2024htj}, the DLA-HEF resummed coefficient function was matched to the NLO coefficient function of Ref.~\cite{Ivanov:2004vd} to provide a uniformly accurate description of the exclusive $J/\psi$ and $\Upsilon$ photoproduction cross sections over a wide range of energies. The DLA-HEF computation is applicable only in the region $\rho=\xi/|x|\ll 1$, and therefore the NLO CF coefficient function provides a more reliable approximation at moderate values of $\rho$. Moreover, one must avoid double counting of the $\rho\ll 1$ asymptotics of the NLO CF coefficient function, which are also contained in the DLA-HEF result. In Ref.~\cite{Flett:2024htj}, this double counting was removed by simply subtracting the $\rho\ll 1$ asymptotics of the NLO CF coefficient function, an approach referred to as \textit{subtractive matching}. In the present paper, both the LO and the resummed coefficient functions are supplemented by their corresponding $O(v^2)$ corrections. The extension of the subtractive matching formula of Ref.~\cite{Flett:2024htj} to include relativistic corrections therefore takes the form
\begin{eqnarray}
    && C_i^{\text{(match.,$S$)}}(x,\xi) \nonumber \\
    && = \delta_{i,g}\Big( C_g^{(0,S)} +  C_g^{(v^2,S)} \Big)(x,\xi) + C_i^{(1,S)}(x,\xi)  \nonumber \\
    && +\Big[ \Check{C}^{(\text{HEF},v^0,S)}_i(\xi/|x|) - \frac{|x|}{\xi}\Check{C}^{(\text{HEF},v^0,S)}_i(1) \nonumber \\
    && + \langle v^2 \rangle \Check{C}^{(\text{HEF},v^2,S)}_i(\xi/|x|)  \Big]\theta(|x|>\xi), \label{eq:sub-match}
\end{eqnarray}
where $i=g,q$, and the dependence on the scales $\mu_F$ and $\mu_R$ has been suppressed to avoid cluttering the notation. Here, $C_g^{(0,S)}(x,\xi)$ is given by eqn.~(\ref{eq:Cg-LO}), $C_g^{(v^2,S)}(x,\xi)$ by eqn.~(\ref{eq:v2-C-LO-exact}), $C_i^{(1,S)}(x,\xi)$ is the NLO coefficient function of Ref.~\cite{Ivanov:2004vd}, which we do not reproduce here for brevity, $\Check{C}^{(\text{HEF},v^0,S)}_i$ is given by eqn.~(\ref{eq:CHEF-v0-rho-closed-form}), and $\Check{C}^{(\text{HEF},v^2,S)}_i$ by eqn.~(\ref{eq:CHEF-deriv-form}). The above-mentioned subtraction of the double counting between the DLA-HEF and NLO CF coefficient functions is implemented by the term $-\frac{|x|}{\xi}\Check{C}^{(\text{HEF},v^0,S)}_i(1)$ in the square brackets in eqn.~(\ref{eq:sub-match}). Indeed, as can be seen from eqn.~(\ref{eq:Crho-v0-exp-as}), at $\rho=1$ only the NLO term in $\alpha_s$ survives in $\Check{C}^{(\text{HEF},v^0,S)}_i$, which, by construction of the resummation, is identical to the $\xi\ll |x|$ asymptotics of $C_i^{(1,S)}(x,\xi)$. No analogous subtraction is required for the $O(v^2)$ correction to the resummed coefficient function because, on the fixed-order side, only the $O(v^2)$ correction to the LO coefficient function is currently available, and therefore no double counting occurs with $\Check{C}^{(\text{HEF},v^2,S)}_i$. Double counting would arise only after including the fixed-order $O(\alpha_s v^2)$ correction (i.e. the $O(v^2)$ correction to the NLO coefficient function), but this correction has not yet been computed for exclusive photoproduction.

\begin{figure}
    \centering
    \includegraphics[width=0.9\linewidth]{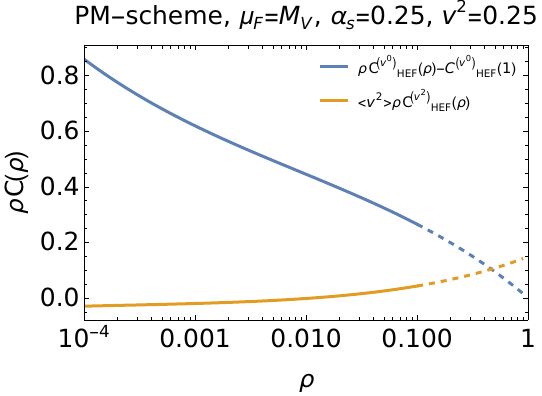}
    \caption{Plots of $\big(\rho \Check{C}_i^{\text{(HEF, $v^0$, PM)}}(\rho,M_V)-\Check{C}_i^{\text{(HEF, $v^0$, PM)}}(1,M_V)\big)/C_{\text{norm.}}$~and~$\rho\Check{C}_i^{\text{(HEF, $v^2$, PM)}}(\rho,M_V)$ $\times\langle v^2 \rangle/C_{\text{norm.}}$, with $C_{\text{norm.}}=-i\pi c^{\text{(PM)}}/2$, for $\alpha_s=\alpha_s(M_{J/\psi})$ and $\langle v^2 \rangle=0.25$, representative of exclusive $J/\psi$ production. The potentially unphysical contribution of the resummation in the region of moderate values of $\rho$ is plotted by the dashed lines.}
    \label{fig:CHEF-plots}
\end{figure}

In Fig.~\ref{fig:CHEF-plots}, the contributions of the first two terms (which we collectively denote as the $O(v^0)$ resummation terms) and of the third term (the $O(v^2)$ resummation term) in the square brackets of eqn.~(\ref{eq:sub-match}) are plotted separately for realistic values of $\alpha_s$ and $\langle v^2 \rangle$, taking $\mu_F=M_{J/\psi}$, corresponding to exclusive $J/\psi$ photoproduction. As expected from eqn.~(\ref{eq:CHEF-PM-v2-asy}), the contribution of the $O(v^2)$ resummation term is numerically negligible compared with the $O(v^0)$ resummation term for $\rho\lesssim 0.1$. Nevertheless, the $O(v^2)$ resummation contribution may still be worth including in phenomenological studies because its $\mu_F$ dependence should partially compensate the $\mu_F$ dependence of the $O(v^2)$-corrected LO amplitude, in much the same way as the $O(v^0)$ DLA-HEF contribution partially cancels the $\mu_F$ dependence of the fixed-order $O(v^0)$ contribution, as demonstrated in Ref.~\cite{Flett:2024htj}.

The subtractive matching procedure defined by eqn.~(\ref{eq:sub-match}) is not unique because it combines two approximations to the coefficient function—fixed-order and resummed—one of which is asymptotic and therefore valid only in the limit $\rho\ll1$. Consequently, the contribution of the resummed coefficient functions at moderate values of $\rho$, for example $0.1\lesssim\rho<1$, is not expected to be physically meaningful. These regions of the curves are shown by the dashed lines in Fig.~\ref{fig:CHEF-plots}. On the other hand, the contribution of this region to the convolution integral (\ref{eq:Ampl-CF}) is not necessarily numerically small. Fortunately, alternative matching prescriptions, such as \textit{Inverse Error Weighting} (InEW) matching~\cite{Echevarria:2018qyi}, suppress these unphysical contributions and allow the matching uncertainty to be assessed systematically. The InEW procedure has already been applied successfully to matching between CF and HEF-resummed coefficient functions in the case of inclusive quarkonium production in Refs.~\cite{Lansberg:2021vie,Lansberg:2023kzf}. The application of this procedure to exclusive photoproduction is currently under investigation.

\section{Conclusions and outlook}\label{sec:concl}

In the present paper, the $O(v^2)$ correction to the DLA-HEF resummed coefficient function for exclusive heavy-quarkonium photoproduction has been computed. The result is presented in the Physical-Mass (PM) scheme, in which the coefficient function is expressed in terms of the physical quarkonium mass $M_V$ and the heavy-quark velocity parameter $\langle v^2 \rangle$, thereby minimising the number of free parameters entering the computation. In the PM scheme, the $O(v^2)$ correction is found to be numerically negligible compared with the $O(v^0)$ resummed coefficient function for the case of $J/\psi$ (Fig.~\ref{fig:CHEF-plots}) at the scale $\mu_R=\mu_F=M_{J/\psi}$. The smallness of the PM-scheme correction is a non-trivial dynamical consequence of the impact-factor computation presented in Sec.~\ref{sec:v2-HEF-IF}. In contrast, the $O(v^2)$ correction to the LO coefficient function in $\alpha_s$ (Ref.~\cite{Blask:2025jua} and Sec.~\ref{sec:LO-CF-v2-corr}) is numerically significant and therefore important for phenomenology.

The impact of the corrections computed in the present work on the phenomenology of exclusive charmonium and bottomonium photoproduction off proton and nuclear targets will be investigated in future phenomenological studies.

\begin{acknowledgments}
The author would like to thank Valerio Bertone, Kari  Eskola, Chris Flett, Vadim Guzey, Jean-Philippe Lansberg, Saad Nabeebaccus, Jyotirmoy Roy, Pawel Sznajder and Jakub Wagner for enlightening discussions which led to the present paper. Author also acknowledges the hospitality of IJClab (Orsay) and the University of Jyv\"askyl\"a (JYU Visiting Fellow Programme Grant 2026, 1263/13.00.05.00/2025).

The work of the author is supported by the ISF grant \#910/23 and by Binational Science Foundation grants \#2022132, \#2024818. 

The tools of the \texttt{FeynCalc} framework were used for the diagrammatic computations in this paper~\cite{Mertig:1990an, Shtabovenko:2016sxi, Shtabovenko:2020gxv, Shtabovenko:2023idz}, Feynman diagrams were drawn with \texttt{feynmp}~\cite{Ohl:1995kr}. \texttt{ChatGPT} was used for the linguistic editing and checking of some equations. The author assumes full responsibility for the scientific content of this paper.
\end{acknowledgments}

\appendix

\section{Finite-$v^2$ analysis at the LO in $\alpha_s$}\label{sec:fin-v2-LO}

  It is possible to compute the coefficient function $C_g(x,\xi)$ as an exact function of relative velocity of heavy quarks $v^2$, using the projector (\ref{eq:NRQCD-proj}). 
  To this end one can simply parametrise the relative four-momentum $k^\mu$ of heavy quarks in eqn.~(\ref{eq:def-p-k}) in terms of the velocity components ($v_x$, $v_y$, $v_z$) in the rest frame of the $Q\bar{Q}$-pair as:
  \begin{eqnarray}
      k^\mu = \frac{M_{Q\bar{Q}}}{2}\bigg[ v_z\frac{N_-^\mu - N_+^\mu}{2}  + v_x N_x^\mu + v_y N_y^\mu \bigg],
  \end{eqnarray}
where the relation between light-cone vectors $N^\mu_{\pm}$ in the center-of-momentum frame of the $Q\bar{Q}$ and $\gamma h$ c.m.s. vectors $n_{\pm}^\mu$ is:
 \begin{equation}
    N_+^\mu = \frac{q_-}{M_{Q\bar{Q}}} n_+^\mu,\, \,  N_-^\mu = \frac{M_{Q\bar{Q}}}{q_-} n_-^\mu. \label{eq:LC-vect-boost}
\end{equation}
In the computation, vectors $N_{x,y}^\mu=n_{x,y}^\mu$ can be defined by their scalar products: $N_{x,y}^2=-1$, $N_xN_y=0$ and $N_{x,y}N_\pm=0$.  The $M_{Q\bar{Q}}$ is expressed in terms of velocity as:
  \begin{equation}
      M_{Q\bar{Q}}=\frac{2m_Q}{\sqrt{1-v^2}},
  \end{equation}
where $v^2=v_x^2+v_y^2+v_z^2$.

For the computation of the CF coefficient function, one considers the pQCD process:
\begin{equation}
    \gamma(q) + g(q_1) \to Q(p_Q) + \bar{Q}(p_{\bar{Q}}) + g(q_2),\label{eq:proc-CF-allv2}
\end{equation}
 with four-momenta of incoming ($g(q_1)$) and outgoing ($g(q_2)$) collinear gluons parametrised as:
\begin{eqnarray}
    q_1^\mu = \frac{n_-^\mu}{2} \frac{M_{Q\bar{Q}}^2}{2\xi q^-} (x+\xi), \\
    q_2^\mu = \frac{n_-^\mu}{2} \frac{M_{Q\bar{Q}}^2}{2\xi q^-} (x-\xi).
\end{eqnarray}

Then one computes the CF coefficient function of the process (\ref{eq:proc-CF-allv2}) using the projector (\ref{eq:NRQCD-proj}) and momenta defined above, as function of $v_x$, $v_y$ and $v_z$ and averages it over the directions of velocity to project-out the contribution of the $L=0$ state:
\begin{eqnarray}
    && \hspace{-5mm}C_g(x,\xi,v^2) = \int\limits_0^{2\pi} d\phi \int\limits_0^{\pi} \frac{d\theta\, \sin\theta}{4\pi} C_g(x,\xi,v_x=v\sin\theta \cos\phi, \nonumber \\
    &&v_y=v\sin\theta \sin\phi, v_z=v\cos\theta). 
\end{eqnarray}
The coefficient function at $v^2=0$ agrees with the standard LO NRQCD result~(\ref{eq:Cg-LO}). The ratio of the velocity-dependent coefficient function to the $v=0$ result:
\[
f^{\text{(MM)}}(x,\xi,v^2) = \frac{C^{\text{(MM)}}_g(x,\xi,v^2)}{C^{(v^0,\text{MM})}_g(x,\xi)},
\]
can be written as:
\begin{widetext}
\begin{eqnarray}
 && f^{\text{(MM)}}(x,\xi,v^2) = \frac{1}{2 \gamma ^4 (\gamma +1) v (x-\xi+i\varepsilon ) (\xi +x-i\varepsilon)} \bigg[ \tanh ^{-1}(v) \bigg(\left(2 \gamma ^4+4 \gamma ^3+3 \gamma ^2-3 \gamma -2\right) \xi
   ^2  \nonumber \\
   && +\left(2 \gamma ^2+2 \gamma +1\right) \gamma ^2 x^2\bigg) -\gamma ^2 v \left(\left(3 \gamma
   ^2+3 \gamma +2\right) \xi ^2+\gamma ^2 x^2\right) \bigg],\label{eq:C(x,xi)-all-v2}
\end{eqnarray}
\end{widetext}
where $\gamma=1/\sqrt{1-v^2}$. The expansion of this function up to $O(v^2)$ is:

\begin{equation}
    f^{\text{(MM)}}(x,\xi,v^2) = 1  -\frac{v^2}{3} \frac{(x^2-7\xi^2)}{(x-\xi+i\varepsilon)(x+\xi-i\varepsilon)}+O(v^4),\label{eq:f-v2-O(v2)}
\end{equation}
which agrees with eqns.~(\ref{eq:v2-C-LO-exact}) and (\ref{eq:gMM-LO}). One can not write-down a meaningful all-order in $v^2$ formula like eqn.~(\ref{eq:C(x,xi)-all-v2}) in the PM-scheme, because no simple relation such as eqn.~(\ref{eq:GK-relation}) between $\langle v^2 \rangle$ and masses exists beyond $O(v^2)$. In general, one should understand that only the ``kinematic'' part of higher-order in $v^2$-corrections is resummed by eqn.~(\ref{eq:C(x,xi)-all-v2}) and it misses all the many-body QCD effects, such as colour-octet contributions of NRQCD. 
\begin{figure}
    \centering
    \includegraphics[width=0.9\linewidth]{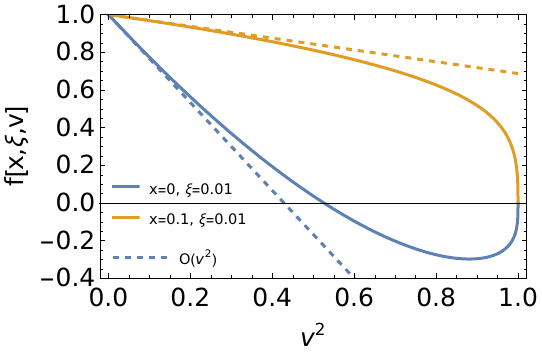}
    \caption{The ratio of the all-order-in-$v^2$ CF coefficient funciton to it's $v^2=0$ limit, eqn.~(\ref{eq:C(x,xi)-all-v2}), as function of $v^2$, taken for $\xi=0.01$ and $x=0$ (blue curve) or $x=0.1$ (orange curve). The $O(v^2)$ expansion (\ref{eq:f-v2-O(v2)}) is plotted by dashed lines.}
    \label{fig:f-v2-plots}
\end{figure}
 The plots of eqn.~(\ref{eq:C(x,xi)-all-v2}) of for several values of $x$  as function of $v^2$ are shown in the Fig.~\ref{fig:f-v2-plots}. In particular this plot makes it clear that for this type of ``kinematic'' $v^2$-corrections, the $O(v^2)$-contribution is very large but it practically exhausts the $v^2$-dependence almost up to $v^2\simeq 0.5$, so the higher-order $v^2$-corrections are essentially negligible for $v^2\lesssim 0.5$.  For the ``dynamical'' $v^2$-corrections, which come from genuine many-body QCD effects, the behaviour may be different, but parametrically they start from $O(v^4)$.

 \section{HEF corrections to the $\delta'(x\pm\xi)$-terms}\label{sec:AppB}


In this appendix, we show how the $\delta'(x\pm\xi)$ terms in eqn.~(\ref{eq:Cg-LO-Im}) arise in the HEF resummation formalism. We also demonstrate that the corresponding higher-order corrections enter at NLL accuracy or beyond and therefore lie outside the accuracy of the present calculation.

The key observation is that the LLA BFKL approximation for the amplitude (\ref{eq:ampl_ga+i-cc+i}) is insensitive, at leading-logarithmic accuracy, to subleading changes in the argument of a large logarithm growing as $\sim \ln(x/\xi)$. Thus, rather than retaining only the strict logarithms of $x/\xi$, one may use the following form of LLA BFKL resummation for the amplitude (\ref{eq:ampl_ga+i-cc+i}):
\begin{eqnarray}
       &&\hspace{-4mm} {\cal M}^{\text{(LLA')}}_{\gamma i} = \int \frac{i\, d^{2-2\epsilon}\T{q}_{T1} d^{2-2\epsilon}\T{ q}_{T2}}{4(2\pi)^{4-2\epsilon}} \int\limits_{-\infty}^{+\infty} dl_+ \frac{{\cal A}_{ab}(l_+,\T{ l}_T,\T{v})}{2\T{ q}_{T1}^2} \nonumber \\
    && \times G\Big(\T{q}_{T1},\T{q}_{T2},\ln\frac{xP^+}{|l^+|}\Big) \int\limits_{-\infty}^{+\infty}dl_- \frac{{\cal B}^{ab}_{i}(l_-,\T{ q}_{T2})}{2\T{ q}_{T2}^2},\label{eq:ampl_ga+i-cc+i-LLA'-00}
\end{eqnarray}
where $G(\T{q}_{T1},\T{q}_{T2},Y)$ denotes the BFKL Green's function, which resums the LLA BFKL series of higher-order corrections $\propto \alpha_s^n Y^n$ and is normalised as
\begin{equation}
    G(\T{q}_{T1},\T{q}_{T2},Y) = \delta^{(2-2\epsilon)}(\T{q}_{T1},\T{q}_{T2})+O(\alpha_s).
\end{equation}
The strict LLA in $\ln(x/\xi)$, used in the main text of the present paper and in Ref.~\cite{Flett:2024htj}, is recovered from eqn.~(\ref{eq:ampl_ga+i-cc+i-LLA'-00}) by taking the $l^+$ residue at the $O(v^0)$ value of $l^+$ given in eqn.~(\ref{eq:lplus-star-v0}):
\begin{equation}
    \ln\frac{xP^+}{|l_{*}^+(\T{q}_{T1})|} = \ln\frac{x}{\xi} - \ln\Big(1+\frac{\T{q}_{T1}^2}{m_Q^2} \Big).\label{eq:ln(rho)+ln(T)}
\end{equation}
The transverse logarithm in this expression does not grow as $x/\xi\to\infty$ and can therefore be neglected at LLA in $\ln(x/\xi)$. The convolution
\begin{equation}
\delta_{ab}\int d^{2-2\epsilon}\T{q}_{T2} G\Big(\T{q}_{T1},\T{q}_{T2},Y\Big) \int\limits_{-\infty}^{+\infty}dl_- \frac{{\cal B}^{ab}_{i}(l_-,\T{ q}_{T2})}{2\T{ q}_{T2}^2}    
\end{equation}
then yields the resummation factor ${\cal C}_{gi}(Y,\T{q}_{T1}^2,\mu_F^2)$ after subtraction of the collinear divergences. The remaining steps follow the procedure described in the main text or in Ref.~\cite{Flett:2024htj}.

Rather than imposing the strict LLA in $\ln(x/\xi)$ at the outset in eqn.~(\ref{eq:ampl_ga+i-cc+i-LLA'-00}), we now perform a more careful expansion in $v^2$. In addition to the explicit $v^2$ dependence of the impact factor ${\cal A}_{ab}$, already included in Sec.~\ref{sec:v2-HEF-IF}, we take into account the $v$ dependence of the positions of the $l^+$ poles of ${\cal A}_{ab}$:
\begin{eqnarray}
    &&\hspace{-4mm} l_{* 1,2}^+(\T{q}_{T1}) = \frac{2}{q_-} \Bigg[ \frac{m_Q^2}{1-v^2} \nonumber \\
    && + \frac{\T{q}_{T1}^2 \mp \dfrac{2m_Q v |\T{q}_{T1}|\sin\theta\cos\phi}{\sqrt{1-v^2}}}{1\pm v\cos\theta} \Bigg], \label{eq:lp12-v^2}
\end{eqnarray}
Here, the angles $\theta$ and $\phi$ parametrise the direction of the vector $\T{v}$ relative to the $\gamma p$ beam axis $Oz$ and the vector $\T{q}_{T1}$. Taking both sources of $v$ dependence into account and averaging over $\theta$ and $\phi$, we obtain the following expansion of the $l^+$ pole contribution to the impact factor ${\cal A}_{ab}$:
\begin{eqnarray}
  &&  \int\frac{d\Omega_{\theta,\phi}}{4\pi} \frac{ \delta^{ab} {\cal A}_{ab}(l_+,\T{q}_{T},\T{v})}{{\bf q}_{T}^2} \propto \delta\left(l^+ - \frac{2(m_Q^2+\T{q}_T^2)}{q_-} \right) \nonumber \\
  && \times \Big(h^{(v^0,S)}(\T{q}_{T}^2) + v^2 h^{(v^2,S)}(\T{q}_{T}^2)  \Big) \nonumber \\
  && + v^2 \bigg[ \tilde{h}_1^{(v^2)}(\T{q}_T^2) \frac{M_V^2}{2q_-} \delta'\bigg(l_+ - \frac{M_V^2+4\T{q}_T^2}{2q_-} \bigg)\label{eq:A(v^2)-delta-expansion} \\
  && + \tilde{h}_2^{(v^2)}(\T{q}_T^2) \frac{M_V^4}{(2q_-)^2} \delta''\bigg(l_+ - \frac{M_V^2+4\T{q}_T^2}{2q_-} \bigg) \bigg]+O(v^4).\nonumber
\end{eqnarray}
The terms involving $\delta'$ and $\delta''$ arise from the expansion of the pole positions in eqn.~(\ref{eq:lp12-v^2}) up to $v^2$. The resulting functions $\tilde{h}_1^{(v^2)}$ and $\tilde{h}_2^{(v^2)}$ are given by
\begin{eqnarray}
    && \tilde{h}_1^{(v^2)}(\T{q}_T^2) = -\frac{2(3M_V^4+6M_V^2\T{q}_T^2 + 8 \T{q}_T^4)}{3(M_V^2+4\T{q}_T^2)^2}, \\
    && \tilde{h}_2^{(v^2)}(\T{q}_T^2) = \frac{8\T{q}_T^2(M_V^2+\T{q}_T^2)}{3M_V^2(M_V^2+4\T{q}_T^2)}. 
\end{eqnarray}
Substituting the expansion (\ref{eq:A(v^2)-delta-expansion}) into eqn.~(\ref{eq:ampl_ga+i-cc+i-LLA'-00}) and expressing the dependence on $m_Q$ in terms of $M_V$ using eqn.~(\ref{eq:GK-rel-mQ}), we obtain the following expression for the resummed coefficient function up to $O(v^2)$:
\begin{eqnarray}
  && \hspace{-4mm} C_i^{(\text{LLA'},S)}(\rho) = \frac{-i\pi c^{(S)}}{2\rho} \int\limits_0^{\infty} d\T{q}_T^2 \int\limits_0^1 dz\; \delta \left(z-\frac{M_V^2}{M_V^2 + 4\T{q}_T^2} \right) \nonumber \\
  && \times \bigg\{ \Big( h^{(v^0,S)}(\T{ q}_T^2) + \langle v^2 \rangle h^{(v^2,S)}(\T{ q}_T^2) \Big) {\cal C}_{gi}\Big(\frac{|\rho|}{z},\T{ q}_T^2,\mu_F^2\Big) \nonumber \\
  && + \langle v^2 \rangle \bigg[ h_1^{(v^2,S)}(\T{q}_T^2) z \frac{\partial}{\partial z} {\cal C}_{gi}\Big(\frac{|\rho|}{z},\T{ q}_T^2,\mu_F^2\Big)\nonumber \\
  && + h_2^{(v^2)}(\T{q}_T^2) \bigg( z^2\frac{\partial^2}{\partial z^2} + 2z \frac{\partial}{\partial z} \bigg){\cal C}_{gi}\Big(\frac{|\rho|}{z},\T{ q}_T^2,\mu_F^2\Big) \bigg] \bigg\},\label{eq:C_LLA'(rho)}
\end{eqnarray}
where:
\begin{eqnarray}
  && \hspace{-13mm} h_1^{(v^2,\text{S})}(\T{q}_T^2) = \frac{M_V^2 (h^{(v^0,\text{S})} + \tilde{h}_1^{(v^2)}) }{M_V^2 + 4\T{q}_T^2} \nonumber \\
  && = - \frac{M_V^2(3M_V^4+16\T{q}_T^4)}{3(M_V^2+4\T{q}_T^2)^3}, \\
  && \hspace{-13mm} h_2^{(v^2)}(\T{q}_T^2) =\frac{M_V^4 \tilde{h}_2^{(v^2)} }{(M_V^2 + 4\T{q}_T^2)^2} = \frac{8M_V^2 \T{q}^2_T(M_V^2+\T{q}_T^2)}{3(M_V^2+4\T{q}_T^2)^3}. 
\end{eqnarray}


As discussed following eqn.~(\ref{eq:ln(rho)+ln(T)}), the deviation of $z$ from unity in eqn.~(\ref{eq:C_LLA'(rho)}) gives rise to an NLL corrections in $\ln(1/\rho)$. The terms involving $\partial/\partial z$ and $\partial^2/\partial z^2$ in eqn.~(\ref{eq:C_LLA'(rho)}) are also beyond LLA: each $z$ derivative reduces the highest power of $\ln(1/\rho)$ at a given order in $\alpha_s$. Consequently, the $O(v^2)$ terms proportional to $h_1^{(v^2,S)}$ and $h_2^{(v^2)}$ contribute only beyond LLA in $\ln(1/\rho)$ and are therefore neglected in the present study.

Finally, to connect this result with the LO in $\alpha_s$, $O(v^2)$ coefficient function in eqn.~(\ref{eq:Cg-LO-Im}), recall that the resummation factor at LO in $\alpha_s$ is (see Sec. 2.3 of Ref.~\cite{Lansberg:2021vie}):
\begin{equation}
    {\cal C}_{gg}(|\rho|,\T{q}_T^2,\mu_F^2) = \delta(\T{q}_T^2) \delta(|\rho|-1)+O(\alpha_s),
\end{equation}

Hence, the $O(\alpha^0_s v^2)$ contribution to eqn.~(\ref{eq:C_LLA'(rho)}) is determined by the functions $h^{(v^2,S)}(\T{ q}_T^2)$ and $h^{(v^2,S)}_1(\T{ q}_T^2)$ in the limit $\T{q}_T^2\to 0$\footnote{The term proportional to $h_2^{(v^2)}$ does not contribute because $h_2^{(v^2)}(0)=0$.}. The term proportional to $h^{(v^2,S)}_1(0)=-1$ contains $\delta'(|\rho|-1)$ and reproduces the terms $\propto \delta'(x\pm \xi)$ in eqn.~(\ref{eq:Cg-LO-Im}). Thus, the (N)NLL terms in the square brackets in eqn.~(\ref{eq:C_LLA'(rho)}) can be viewed as higher-order ``descendants'' of the $\delta'(x\pm \xi)$ terms in eqn.~(\ref{eq:Cg-LO-Im}).

\end{document}